\documentclass[a4paper,11pt]{article}
\usepackage{jinstpub} 
\usepackage{lineno}

\input{ds_macro.def}
\usepackage{orcidlink}

\newcommand{\isotope}[2]{\mbox{$^{#1}$#2}}

\newcommand{\insitu}{\emph{in situ}}

\title{\boldmath Long-term temporal stability of the DarkSide-50 dark matter detector}

\collaboration{The DarkSide-50 Collaboration}

\author[a]{P.~Agnes,}
\author[b]{I.F.M.~Albuquerque,}
\author[c]{T.~Alexander,}
\author[d]{A.K.~Alton,}
\author[b]{M.~Ave,}
\author[c]{H.O.~Back,}
\author[e,f]{G.~Batignani,}
\author[g]{K.~Biery,}
\author[h]{V.~Bocci,}
\author[i]{W.M.~Bonivento,}
% ...

\affiliation[a]{Department of Physics, Royal Holloway University of London, Egham TW20 0EX, UK}
\affiliation[b]{Instituto de F\'isica, Universidade de S\~ao Paulo, S\~ao Paulo 05508-090, Brazil}
\affiliation[c]{Pacific Northwest National Laboratory, Richland, WA 99352, USA}
\affiliation[d]{Physics Department, Augustana University, Sioux Falls, SD 57197, USA}
\affiliation[e]{INFN Pisa, Pisa 56127, Italy}
\affiliation[f]{Physics Department, Universit\`a degli Studi di Pisa, Pisa 56127, Italy}
\affiliation[g]{Fermi National Accelerator Laboratory, Batavia, IL 60510, USA}
\affiliation[h]{INFN Sezione di Roma, Roma 00185, Italy}
\affiliation[i]{INFN Cagliari, Cagliari 09042, Italy}
% ...

\abstract{
The stability of a dark matter detector on the timescale of a few years is a key requirement due to the large exposure needed to achieve a competitive sensitivity.
It is especially crucial to enable the detector to potentially detect any annual event rate modulation, an expected dark matter signature.
In this work, we present the performance history of the \DSf\ dual-phase argon time projection chamber over its almost three-year low-radioactivity argon run.
In particular, we focus on the electroluminescence signal that enables sensitivity to sub-keV energy depositions.
The stability of the electroluminescence yield is found to be better than 0.5\%.
Finally, we show the temporal evolution of the observed event rate around the sub-keV region being consistent to the background prediction.
}

\keywords{Dark Matter detectors (WIMPs, axions, etc.),Time projection Chambers (TPC)}

\arxivnumber{2311.18647} 

\begin{document}
\maketitle
\flushbottom

\section{\label{sec:intro}Introduction}
Dark matter direct detection experiments with an Earth-based detector look for energy depositions from an interaction between dark matter and the detector medium.
In the Standard Halo Model, the orbital motion of the Earth around the Sun produces an annual variation of the relative velocity of the Earth with respect to the Galactic center, and consequently with respect to the static dark matter halo.
As the energy deposition depends on the relative dark matter velocity, the count rate above the detector threshold may show an annual oscillation~\cite{Drukier:1986tm}.
The detection of such an oscillation is a promising avenue towards the discovery of dark matter.

As of today, several experiments have conducted an annual modulation search using a variety of detector technologies.
The DAMA Collaboration (DAMA/NaI and DAMA/LIBRA) has operated a large array of NaI crystal detectors in the deep underground site of \textit{Laboratori Nazionali del Gran Sasso} (LNGS) in Italy, observing a clear modulation~\cite{Bernabei:2013xsa,Bernabei:2021kdo}.
The signal is in the energy window of \SIrange{0.75}{6}{\keVee}~(``electron-recoil equivalent'') and appears to be consistent in phase and period with the dark matter hypothesis.
However, other experiments using liquid xenon detectors have failed to confirm this result~\cite{XENON:2017nik,LUX:2018xvj,XMASS:2022tkr}.
The interpretation of the observed modulation in several dark matter models, including Weakly Interacting Massive Particles (WIMPs), has also been constrained by many experiments, e.g.~\refscite{LUX:2016ggv,SuperCDMS:2017mbc,XENON:2018voc,DarkSide:2018kuk,DEAP:2019yzn,COSINE-100:2021xqn,XMASS:2022tkr,LZ:2022ufs,XENON:2023sxq}.
DAMA's observation represents a long-standing anomaly, while other experiments adopting technologies similar to DAMA's are making progress towards definitively testing the result~\cite{Amare:2021yyu,COSINE-100:2021zqh,COSINE-100:2022dvc}.

Dual-phase noble liquid time projection chambers (TPCs), if employing solely the ionization signal, reach sensitivity to lower-energy depositions, far below DAMA's threshold~\cite{DarkSide-50:2022qzh,XENON:2016jmt,XENON:2019gfn}.
However, the long-term operation of such detectors requires active cryogenic controls to maintain a high-purity target.
Thus, careful attention has to be paid to the detector performance, as the stability of the event rate is highly sensitive to the cryogenic conditions.

The \DSf\ experiment performed a direct dark matter search using a liquid argon TPC at LNGS~\cite{DarkSide:2015cqb,DarkSide:2018kuk,DarkSide:2018bpj,DarkSide:2018ppu,DarkSide-50:2022qzh,DarkSide:2022dhx,DarkSide:2022knj}.
Here, we present a study of the stability of the \DSf\ TPC performance over the data-taking period of 2.5 years.
We also present the temporal evolution of both the expected and observed event rate in the low energy region that is of particular interest for the dark matter search.

\section{\label{sec:detector}The \DSf\ detector}
The \DSf\ TPC measures scintillation (S1) and ionization signals from an energy deposition in the liquid phase.
It is filled by an active mass of \SI{46.4\pm0.7}{\kilogram} of low-radioactivity argon extracted from a deep underground source (UAr)~\cite{Back:2012pg,DarkSide:2012fps,Xu:2012ffp,DarkSide:2015cqb}.
A thin gas layer (``gas pocket'') lies above the active volume to convert the ionization signal to electroluminesence light (S2).
Two arrays of nineteen 3-inch Hamamatsu R11065 photomultiplier tubes (PMTs) are located at the top and the bottom of the TPC.
As the argon light emission lies in the vacuum ultraviolet (\SI{128}{\nano\meter}), it is downshifted to the visible spectrum before reaching the PMT by a wavelength shifter (TPB, tetraphenyl butadiene) coated on all inner surfaces facing the fiducial volume.
The PMT signals are routed to a digitizer that triggers upon a coincidence of at least two PMTs above \SI{0.6}{photoelectron} (PE) within \SI{100}{\nano\second}~\cite{DarkSide:2017odo}.
The average S2 PE per ionization electron (\(g_2\)) in the inner \(\sim\)\SI{23}{\centi\meter}-diameter cylindrical volume (\(\sim\)\SI{20}{\kilogram}), which defines the fiducial volume for the S2-only analysis in~\refscite{DarkSide:2018bpj,DarkSide:2018ppu,DarkSide-50:2022qzh,DarkSide:2022dhx,DarkSide:2022knj}, is measured to be \SIrange{17}{23}{\pe\per\el}, depending on the radial position of an electron extracted to the gas phase.
The resolution of a single electron signal is \SI{0.27}{\el}.
The trigger efficiency for S2 from that volume reaches \(\sim\)100\% at \SI{1.3}{\el}~\cite{DarkSide:2018bpj}.

\begin{figure}[t]
    \centering
    \includegraphics[width=0.65\linewidth]{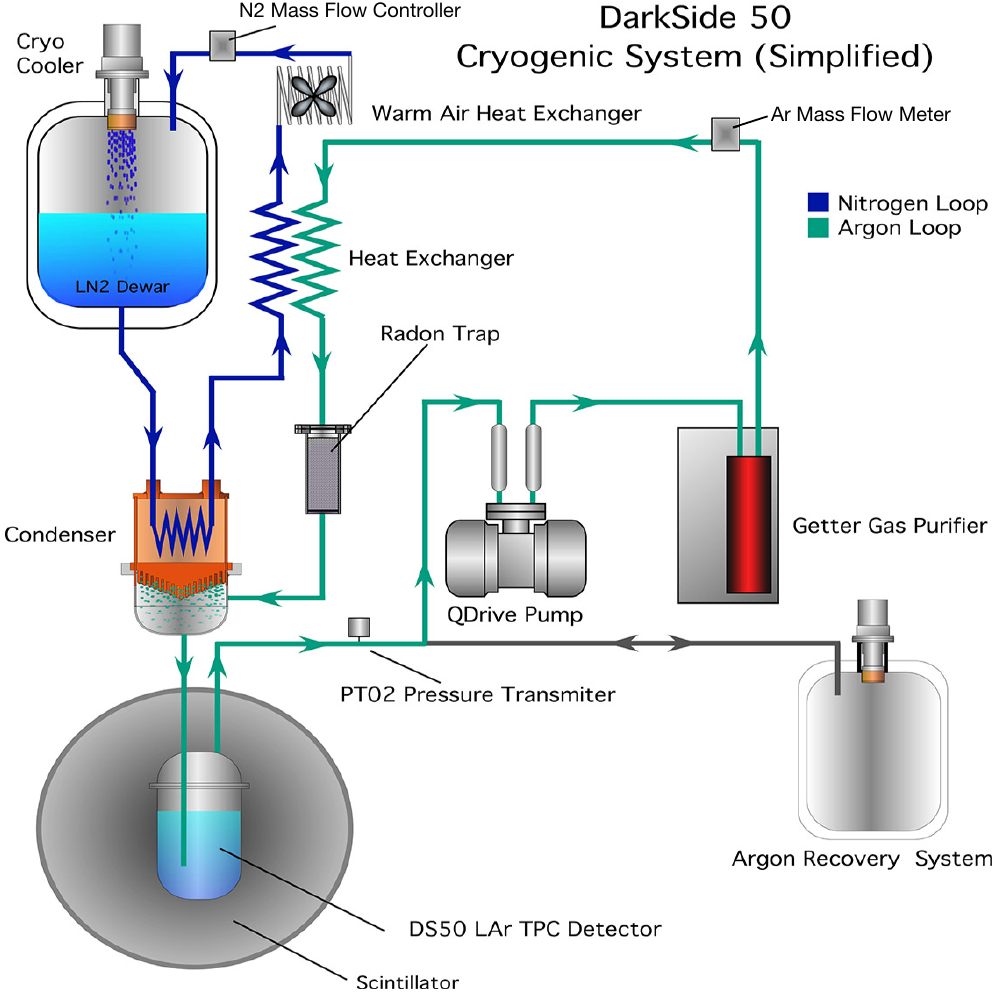}
    \caption{Schematic overview of the \DSf\ cryogenic system.}
    \label{fig:system}
\end{figure}

The argon inside the TPC is handled by the system shown in~\reffig{fig:system}. 
Gaseous argon extracted from the cryostat containing the TPC passes through a commercial getter gas purifier (SAES Monotorr PS4-MT50-R-2~\cite{saes}) where electronegative and VUV light absorbing impurities, such as \ce{O2}, \ce{H2O}, \ce{CH4}, and \ce{N2}, are reduced to parts per billion levels.
The gas is then pre-cooled by a heat exchanger coupled to cold nitrogen and is passed through a charcoal trap for radon removal.
A condenser following the filters returns purified liquid argon directly into the TPC.
The gas pocket of the TPC is maintained by a ``boiler'' on the side of the TPC which extracts liquid argon from the inside of the TPC and returns boiled-off gas.
The electronics and the filters are located in the clean room on top of the water and liquid scintillator tanks surrounding the cryostat.

\begin{figure*}
    \begin{minipage}[t]{0.49\hsize}
        \centering
        \includegraphics[width=\linewidth]{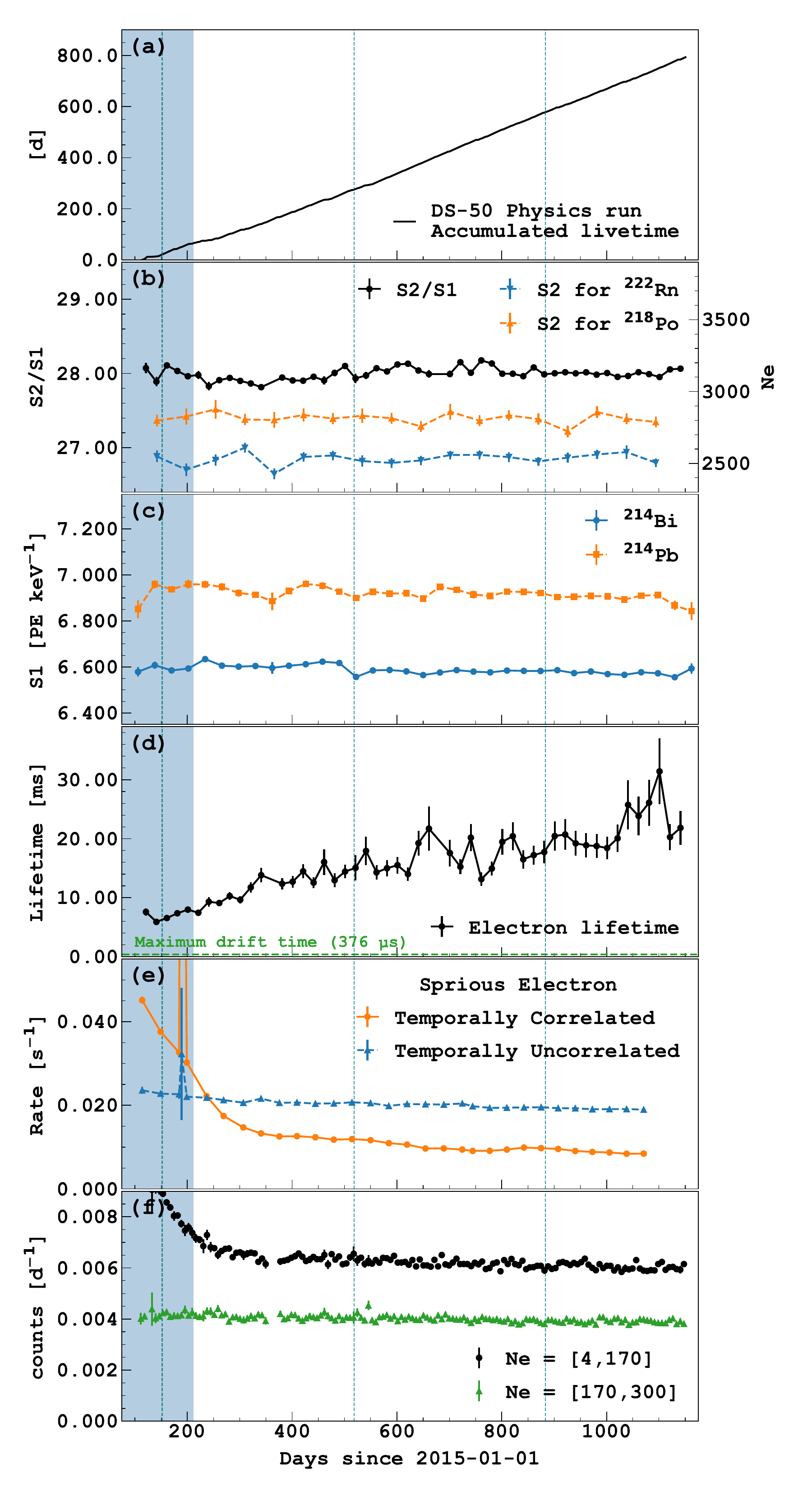}
    \end{minipage}    
    \begin{minipage}[t]{0.49\hsize}
        \centering
        \includegraphics[width=\linewidth]{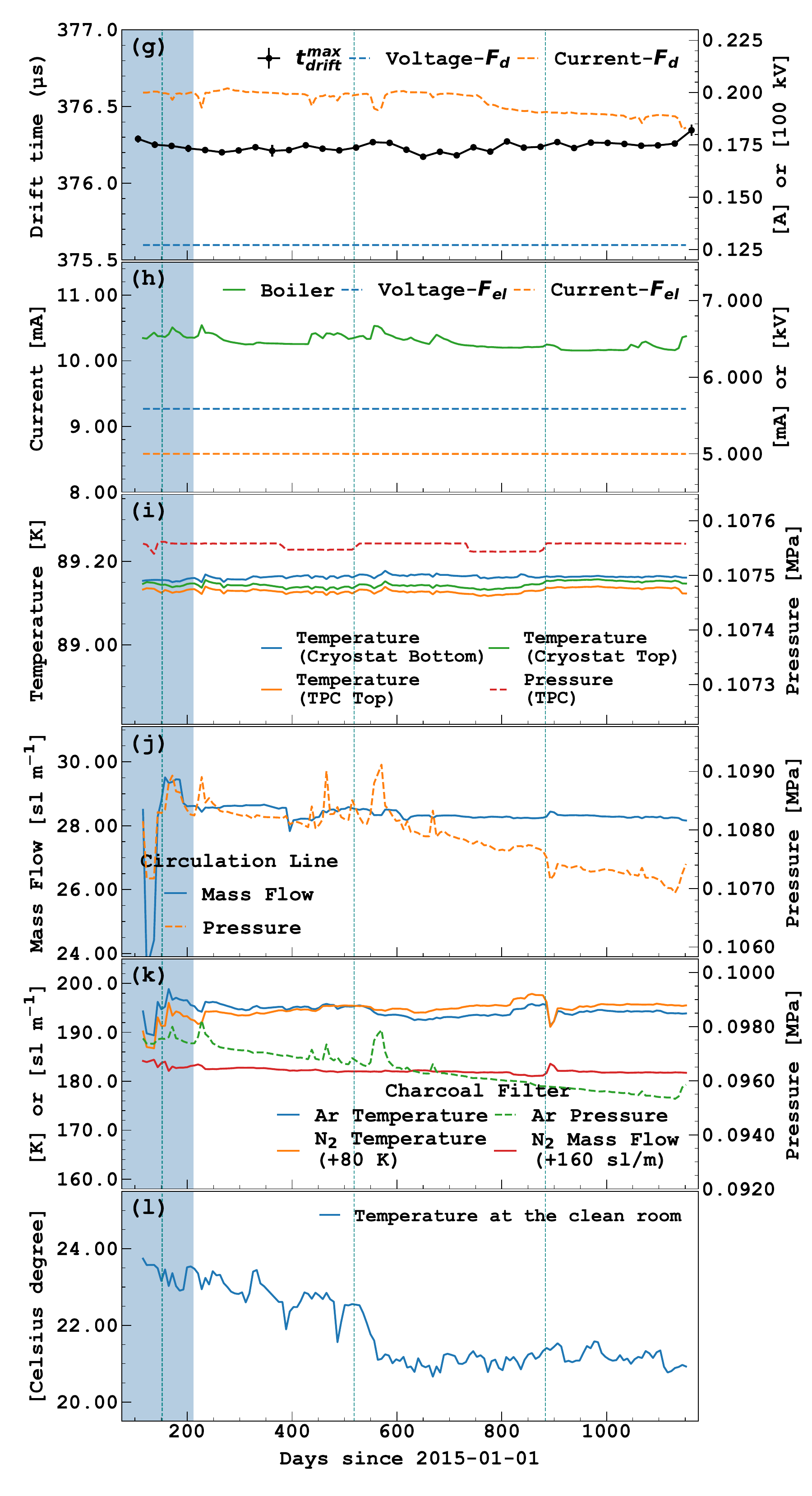}
    \end{minipage}    
    \caption{Temporal evolution of the detector parameters of interest for this analysis. 
    \textbf{a:} accumulated livetime over the time,
    \textbf{b--d:} the parameters measured by \(\beta\)- or \gr\ events from the TPC such as the S2/S1 ratio, S1 detection efficiency, and electron lifetime.
    \textbf{e:} temporally correlated and uncorrelated SE rates.
    \textbf{f:} the observed event rates of both the RoI and higher energy region.
    \textbf{g:} full drift time of the TPC measured by the event near the cathode and the measured current at the voltage supplier.
    \textbf{h--l:} the parameters monitored by sensors inside the system such as temperatures, pressures, and gas flow rate.
    The blue-shaded period represents the period devoted to the \isotope{37}{Ar} calibration.
    The vertical dashed lines represent June 2nd of each year when the dark matter induced event rate is expected to be maximum.}
    \label{fig:slow}
\end{figure*}

Data collection lasted 35 months, between April 2015 when the detector was filled with UAr, until February 2018 (see \reffig{fig:slow}(a)).
The UAr was extracted in southwestern Colorado, purified at Fermi National Accelerator Laboratory, and transported to LNGS~\cite{Back:2012pg,DarkSide:2012fps}.
The data from the first four months yielded the initial dark matter search~\cite{DarkSide:2015cqb} and were also used to calibrate the detector thanks to the presence of a measurable amount of \isotope{37}{Ar} (the half-life of \SI{35.0}{\day}), a product of cosmogenic activation of argon during its transportation~\cite{DarkSide:2021bnz}.
The last 27 months of data taking were used for improved dark-matter searches, see~\refscite{DarkSide-50:2022qzh,DarkSide:2022dhx,DarkSide:2022knj}.
The modulation analyses, presented in this paper as well as in \refcite{DarkSide-50:2023fgf}, use all the data except the initial four months', with the total livetime of \SI{693.3}{\day}.

More details on the \DSf\ apparatus can be found in~\refscite{DarkSide:2014llq,DarkSide:2017odo}.

\section{\label{sec:stability}Stability of the detector performance}

Variations in the performance of the experimental system may affect the detector response and introduce calibration uncertainties or artificial event rate modulations.
From its accumulated exposure and measured radioactive backgrounds, \DSf\ can be sensitive to a modulation amplitude of \(\sim\)1\% of the observed event rate.
The sensitivity would decrease if the experimental system exhibits any variations larger than that level.

Three parameters especially influence the detector response.
One is the electric field in the fiducial volume, \(F_d\), which at the nominal value of \SI{200}{\volt\per\centi\meter} affects the scintillation and ionization yields approximately linearly~\cite{Doke:2002oab}.
The other two are the average numbers of detected PE per scintillation photon, \(g_1\), and per ionization electron, \(g_2\).
While \(g_1\) impacts the energy reconstruction in high-mass (\(>\)\SI{10}{\giga\eV\per\square\c}) dark matter searches~\cite{DarkSide:2014llq,DarkSide:2015cqb,DarkSide:2018kuk}, \(g_2\) is critical for the S2-only analyses.
Indeed, a toy Monte Carlo simulation study, which accounts for the expected background model (see \refsec{subsec:bgmodel}) and the detector response model, suggests that a 1\% fluctuation of \(g_2\) could lead to a non-physical \(\mathcal{O}(1)~\mathrm{keV}\) signal.

Throughout the data-taking period we continuously monitored these and other environmental and detector parameters using source calibration data and an array of temperature, pressure and flow sensors in the argon system and monitoring the stability of voltage delivery to the TPC.
The possible impact of the monitored parameters on the scintillation and ionization observables is described in the following section.

\subsection{\label{subsec:methods}Methodology}
Besides the aforementioned parameters \(F_d\), \(g_1\), and \(g_2\), other parameters having a clear correlation with the light signals are the gain and resolution of the PMTs to a single photoelectron, the electroluminescence field \(F_{g}\), the temperature and pressure of the gas pocket of the TPC, and the liquid argon purity.
Their potential impact is evaluated by propagating the observed fluctuation according to the expected correlation.

On the other hand, for other parameters the correlation with the observable is not known \textit{a priori}.
For such parameters, we calculate different correlation coefficients, namely Pearson, Kendall and Spearman, to the event rate that assess either linear or rank correlation.
Here, we use the residual of the background-only fit, discussed in \refsec{sec:result}, instead of the raw event rate, such that the decay of known short-lived isotopes is accounted for.
Furthermore, a fluctuation of one parameter could produce a delayed effect in time on the event rate.
In order to catch such a time-delayed correlation, we repeat the correlation coefficient calculation by introducing a time shift ranging from 1~day to 2~months.



Moreover, the Lomb-Scargle (LS) periodogram~\cite{Lomb:1976,Scargle:1982} is supplementarily used to look for any periodical fluctuation of the parameters.
The output power spectrum is compared to the false alarm probability calculated with the Bootstrap method~\cite{2018ApJS..236...16V}, so that the significance of the periodic fluctuation in a particular period is quantified, as described in~\refsec{subsec:other_slc}.



\subsection{\label{subsec:history}Stability of each parameter}

\subsubsection{\label{subsec:system}Cryogenic system}
The liquid argon inside the cryostat is maintained by a proportional-integral-derivative (PID) controller which keeps the pressure at \SI{1.08E5}{\Pa}.
The controller adjusts the cooling power from the nitrogen loop by the mass flow controller shown in \reffig{fig:system}.
Another PID loop controls the temperature of the cryocooler to condense the nitrogen.
The whole system is monitored by \(\sim\)70 sensors measuring temperature, pressure, gas flow, and heating power at various points of the system.

The time profile of the argon pressure and the cryostat temperatures are shown in \reffig{fig:slow}(i).
The pressure is used in the PID control algorithm and is stable to better than \SI{35}{\Pa}.
Temperatures are stable to within \SI{\pm0.02}{\kelvin}.
No visible impact is anticipated in terms of \(g_2\), as will be discussed in \refsec{subsec:g2}.

Figures~\ref{fig:slow}(j-k) show the parameters associated with the argon circulation line.
The mass flow of the circulation line is kept between \SIrange{28}{29}{sl\per\minute}, while the pressure inside the line decreases continuously through the data-taking period by 1.5\%.
This change reflects the change in the gas temperature inside the loop.
The instabilities of the temperatures and pressures related to the radon trapping part are around 1\%, which fluctuate coherently.
There is an indication that the spurious electron (SE) rate may have a correlation with the temperatures at the radon trap, the object of a paper in preparation.
However, we do not find any way for these instabilities to affect the ionization signal above the typical energy threshold of \SI{4}{\el}.
Indeed, any of the three aforementioned correlation coefficients between these parameters and the event rate for different \(N_e\) ranges above \SI{4}{\el} is less than 0.06 (p-value larger than 0.15).

\subsubsection{\label{subsec:pmt}PMT response}
The characterization of the PMTs is performed roughly every \SI{12}{\hour} by illuminating the TPC with a blue laser~\cite{DarkSide:2014llq} so that the gain changes are calibrated out in such timescale.
All PMTs show a similar trend in time, \textit{i.e.} a slight monotonic decrease in gain by \(\sim\)5\% over the data-taking period, while the single photoelectron resolution remains constant.
The fluctuations of both the gain and the resolution are measured to be \(\sim\)1\% by looking at the distribution of these parameters.
The impact of PMT instabilities and temporal gain changes on the results of dark matter searches was assessed, with a special focus on their effect on trigger and event selection efficiency.
The study demonstrates that this level of fluctuations has no visible effect on the results.

\subsubsection{\label{subsec:efield}Electric fields}
The stability of \(F_d\) is traced \insitu\ by the drift time of events at the very bottom of the TPC, \(t_\mathrm{drift}^\mathrm{max}\), where the drift time is defined as the time difference between S1 and S2.
Its instability is measured to be \(\mathcal{O}(0.01\%)\), as shown in~\reffig{fig:slow}(g), too small to affect the detector response.
The measured voltage at the power supply providing the appropriate potential to each electrode shows fluctuations \(<\)0.01\%.
On the other hand, the current at the power supplier undergoes a gradual change as big as \(\sim\)10\%.
Although the cause of this variation remains unclear, a small correlation coefficient, of less than 0.05, to the observed event rate is found with the Pearson method.
Therefore, we exclude that a modulation search could be affected by such a variation.

Since the distance between the gate grid and the anode does not change over time, \(F_{g}\) depends on the applied potential difference between the gate grid and the anode, and the height of the gas pocket.
\reffiginitpar{fig:slow}(h) shows the related parameters.
The high voltage supplied to achieve the potential difference is stable during the data-taking period.
A power supply driving the boiler to maintain the gas pocket fluctuates by \(\pm\)1\%.
This however does not directly affect the gas pocket condition since it is set by a hole on the side wall of the TPC through which excess gas bubbles out to the cyrostat.
The stability of \(g_2\) is independently studied in~\refsec{subsec:g2}.

\subsubsection{\label{subsec:g1}Scintillation light yield}
The temporal variation of \(g_1\) is traced by mono-energetic peaks from background \grs.
The peak positions of the \SI{352}{\keV} \gr\ from \isotope{214}{Pb} and \SIlist{609}{\keV} \gr\ from \isotope{214}{Bi} are shown in \reffig{fig:slow}(c). 
Fluctuations of \(g_1\) over the data-taking period are evaluated as 0.3\%.
It is also worth noting that the fluctuations from these two peaks do not appear to be correlated.

An independent test of the \(g_1\) stability was performed by looking at calibration campaigns injecting \isotope{83m}{Kr} diffused radioactive source into the TPC.
The fluctuation between the three campaigns performed during the period is \(\sim\)0.4\%~\cite{DarkSide:2018kuk}.

\subsubsection{\label{subsec:g2}Electroluminescence yield}
As mentioned, since \(g_2\) affects directly the observed ionization spectrum, its stability is of particular interest.
It is monitored via the S2/S1 ratio, \(R\), of the background \br\ events whose energy is higher than the region of interest (RoI) for dark matter searches (\SIrange{3}{170}{\el}).
The parameter \(R\) is corrected by the electron lifetime \(\tau_e\) as obtained in~\refsec{subsec:purity}.
\reffiginitpar{fig:slow}(b) shows the temporal evolution of \(R\), showing a fluctuation of~0.4\%.
Taking into account the fluctuation of \(g_1\), that of \(g_2\) is evaluated to be no more than~0.5\%.

An auxiliary analysis traces the S2 spectrum from \ar\ events that happened inside the bulk UAr.
By using the S1 yields to select \isotope{222}{Rn} and \isotope{218}{Po} events~\cite{DarkSide:2016ddo}, we get the monthly averaged S2 yield for each of them.
The instability is less than 1.5\% as shown in~\reffig{fig:slow}(b), where the sensitivity is limited by the statistical uncertainty.

It is known that the electroluminescence yield has a positive linear relationship with the electric field and a negative linear relationship with the number density of argon atoms.
As a cross check of our measurements, based on~\refcite{Monteiro:2008zz}, we calculate the fluctuation by propagating the measured fluctuations of the temperature and pressure inside the TPC.
It predicts a negligible (relative change of \(<\)\(\mathcal{O}(10^{-4})\)) fluctuation of \(g_2\) as consistent with the observation above.

\subsubsection{\label{subsec:purity}Liquid argon purity}
Any impurities inside liquid argon may cause the deterioration of the detector's performance.
In particular, electronegative impurities, such as \ce{O2}, \ce{H2O}, and \ce{CH4}, absorb drifting electrons during their path to the gaseous phase and more heavily suppress electrons closer to the cathode.
The electron lifetime \(\tau_e\) is measured by looking at a dependence of \(R\) on the path length of the electron.
\reffiginitpar{fig:slow}(d) shows the temporal evolution of \(\tau_e\), which increases from \SI{5}{\milli\second} (corresponding to an \ce{O2} equivalent concentration of \SI{60}{\ppt}~\cite{WArP:2008dyo}) to \(>\)\SI{20}{\milli\second} (\SI{15}{\ppt}), while its fluctuation is \(\sim\)\SI{1}{\milli\second}.
In addition, \(\tau_e\) is well over an order of magnitude longer than the TPC full drift time of \SI{376}{\micro\second}.
Such a long lifetime and a small fluctuation with respect to the drift time are expected to have no significant impact on the observed event rate, as confirmed by a Monte Carlo simulation incorporating the suppression probability of the drift electrons in that range.

\subsubsection{\label{subsec:other_slc}Overall parameters stability}
As mentioned in \refsec{subsec:methods}, for most of the parameters there is no expected \textit{a priori} mechanism responsible for causing sizable fluctuations of the event rate.
Nevertheless, we performed an analysis on all sensors to assess a potential seasonal change and a non-trivial influence they may have.
All of the aforementioned three correlation coefficients are well contained, between $-0.07$ and $0.10$ (p-values larger than $0.01$), for different \(N_e\) ranges relevant to the dark-matter search.
In addition, as anticipated the time-delay analysis returned low correlation coefficient values, between $-0.08$ and $0.10$, showing no sign of potential delayed impact in the TPC event rate.
Owing to the overall stability of these parameters as well as the low correlation they have with the TPC event rate, we can safely affirm that none of the small visible fluctuations are impactful.

Finally, we perform a sanity analysis on all of the parameters based on the LS periodogram.
We find that only a few parameters have a periodicity of \SI{1}{\year} with a \(3\sigma\) significance.
Namely, the argon pressure and temperature inside the cryostat (\reffig{fig:slow}(i)), the liquid level of the nitrogen dewar, and temperatures of auxiliary pumps inside the system.
However, thanks to the already presented stability of the argon pressure and temperature, as well as the aforementioned correlation analysis yielding coefficients lower than, 0.03, 0.07, -0.05, -0.07, respectively, we consider that the few parameters with periodicity found with the LS analysis have no impact on the final result.
We also note that we do not find any known mechanism of affecting the TPC performance for the latter two parameters.


\section{\label{sec:result}TPC event rate stability}
\subsection{\label{subsec:bgmodel}Expectation from radioactivity}
\begin{figure}[t]
    \centering
    \includegraphics[width=0.65\linewidth]{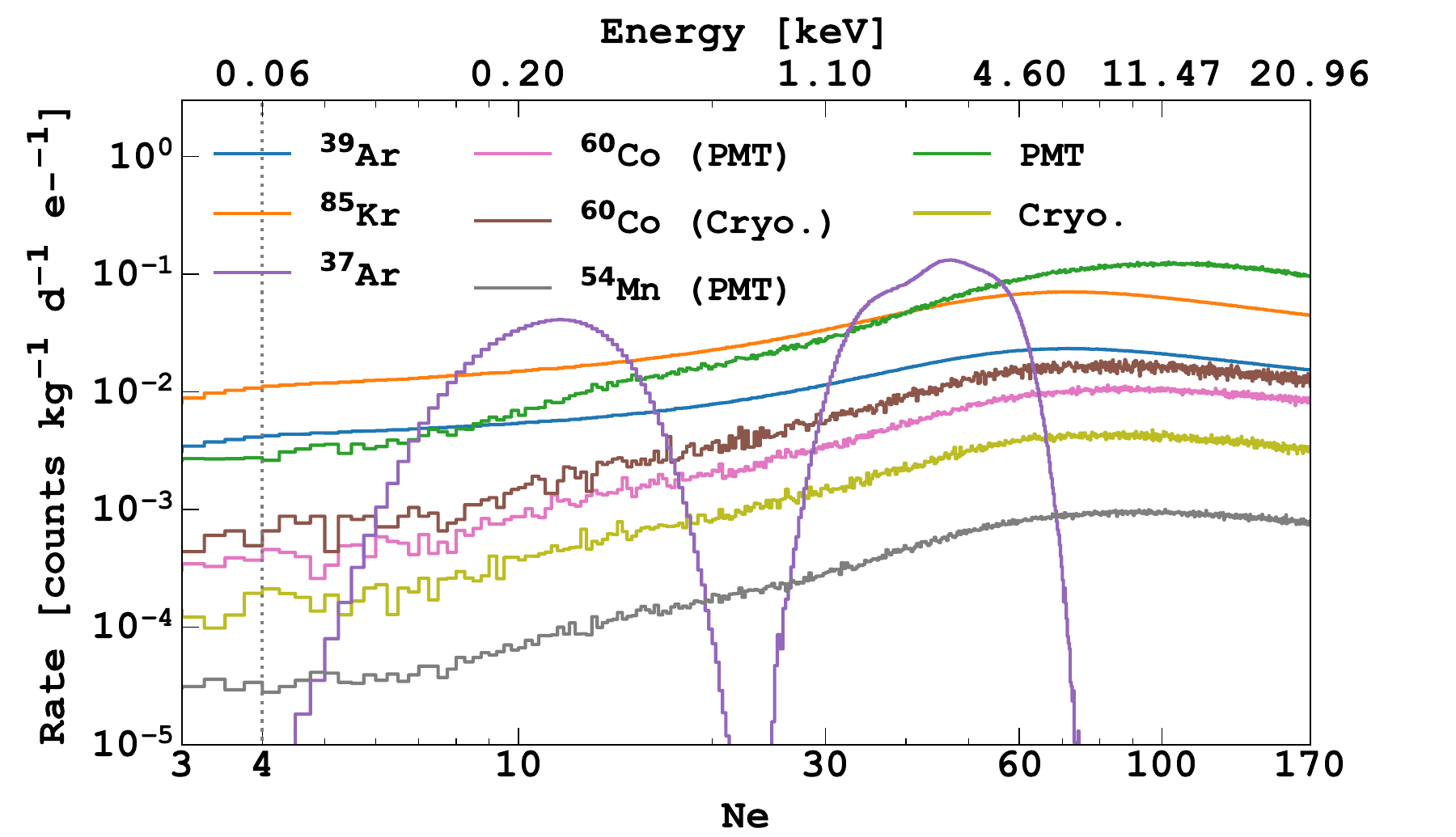}
    \includegraphics[width=0.65\linewidth]{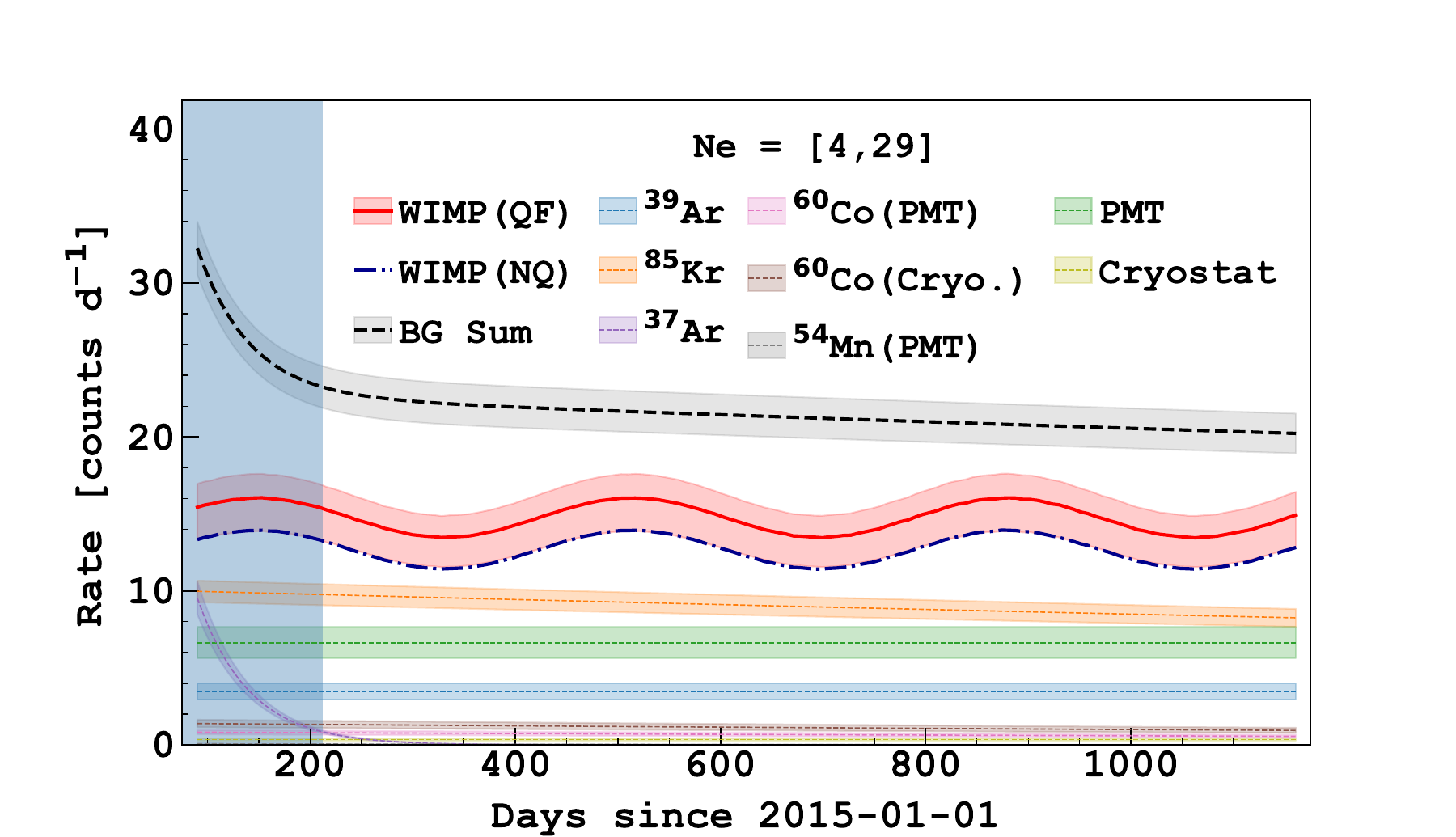}    
    \caption{Top: background model of each component with their total uncertainties including both shape and amplitude systematics. The amplitude of each component shown here is normalized at \SI{123}{\day} passed since the reference day.
    Bottom: temporal evolution of the expected rate from each background source within \SIrange{4}{29}{\el}. 
    Also shown are that of WIMP with \SI{3}{\GeV\per\square\c} assuming WIMP-nucleon cross-section equal to \SI{3E-41}{\square\centi\meter} with (QF) and without (NQ) quenching fluctuations (see text for more detail).
    The blue-shaded period is same as in~\reffig{fig:slow}.}
    \label{fig:bg_ne}
\end{figure}

The majority of the observed events in the RoI come from diffused \(\beta\)-emitters (\isotope{39}{Ar} and \isotope{85}{Kr}) and \(\gamma\)- and x-rays from radioactive contamination in the PMTs and the cryostat.
The decay of \isotope{37}{Ar} also makes a significant contribution until it decays away.

Contributions to the ionization spectrum from each component are evaluated, as done in~\refcite{DarkSide-50:2022qzh} and shown in~\reffig{fig:bg_ne}~(top).
Here, we briefly summarize the way we evaluate them.
For the \(\beta\)-emitters, the shape of the energy spectra is derived from the theoretical calculation of the energy transition of the \(\beta\)-decay, while its activity is evaluated in the data itself in a higher-energy sideband.
We then apply the detector response to the spectra.
For the x- and \grs\ from the detector materials, we perform a \texttt{Geant4}-based Monte Carlo simulation~\cite{DarkSide:2017wdu} to get the ionization spectrum, while the total normalization and associated uncertainty are determined by the result of the material screening campaign, which happened before the detector was commissioned.
The only isotope not discussed in~\refcite{DarkSide-50:2022qzh} is \isotope{37}{Ar}. 
In this case, the spectrum shape is derived by simulating the detector response to the monochromatic energy depositions of \SI{2.83}{\keV} (from the decay via K-shell electron capture) and \SI{0.277}{\keV} (L1-shell), following the procedure in~\refcite{DarkSide:2021bnz}.
The activity is directly measured by the temporal evolution of the event rate in the low energy region for the first four months of data to be \SI{0.42\pm0.03}{\milli\becquerel\per\kilogram}.
This measurement is found to be consistent with that calculated from the argon activation history using nuclear data libraries within~\(\sim\)1\(\sigma\)\footnote{Paper in preparation.}.
\reffiginitpar{fig:bg_ne}~(bottom) shows the expected temporal evolution of these background events.
In addition to \isotope{37}{Ar}, \isotope{60}{Co} (\CoSixZeroHalfLife) and \isotope{85}{Kr} (\KrEightFiveHalfLife) have lifetimes compatible to the data taking period.

 


\subsection{\label{subsec:bgfit}Analysis with the background model}
\begin{figure}[t]
    \centering
    \includegraphics[width=0.65\linewidth]{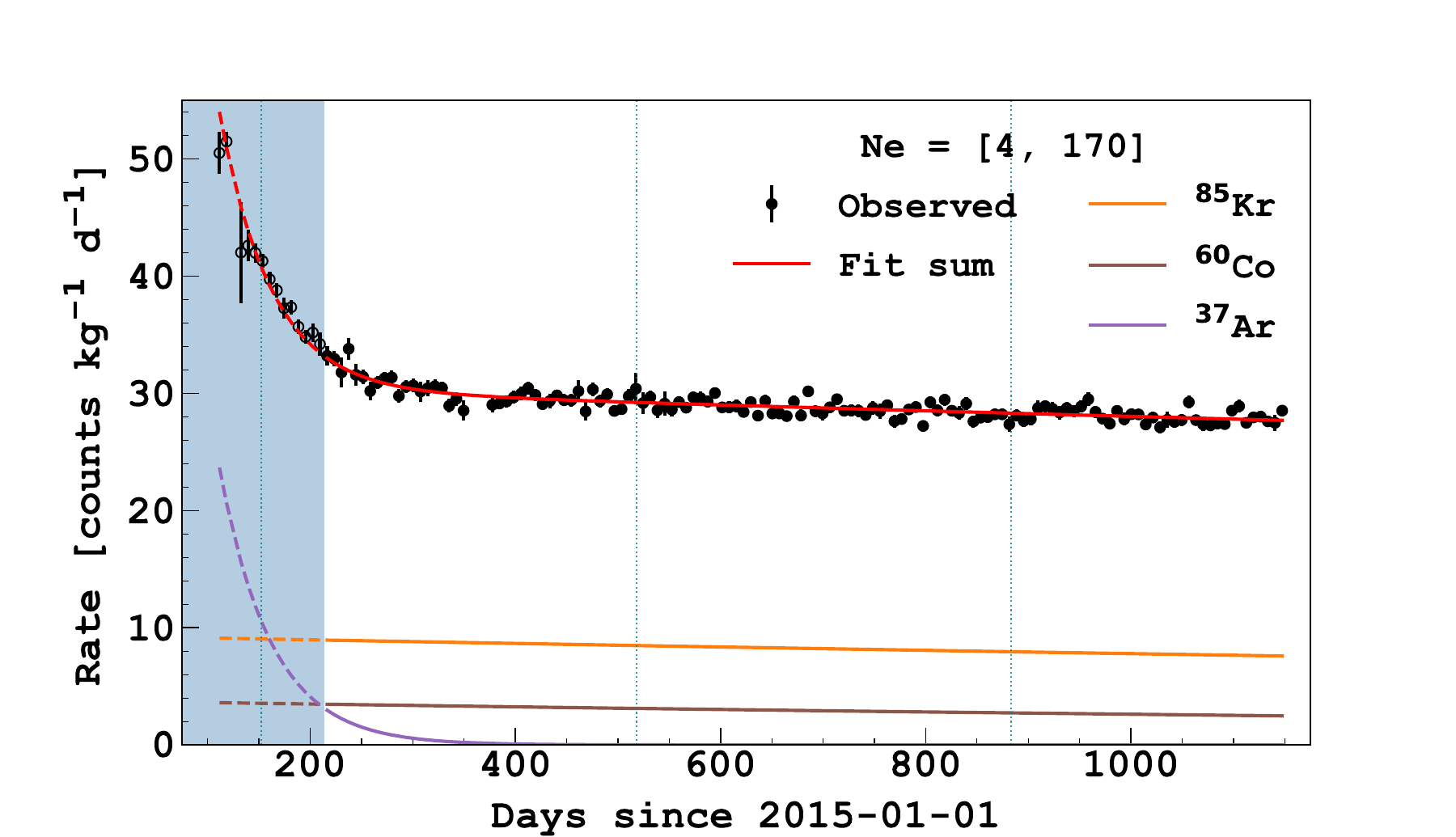}
    \caption{Observed event rate in~\SIrange{4}{170}{\el} and the fit with the background model.
    The blue-shaded period is the same as in~\reffig{fig:slow} and is excluded from this analysis.}
    \label{fig:bgfit}
\end{figure}

\begin{figure}[t]
    \centering
    \includegraphics[width=0.65\linewidth]{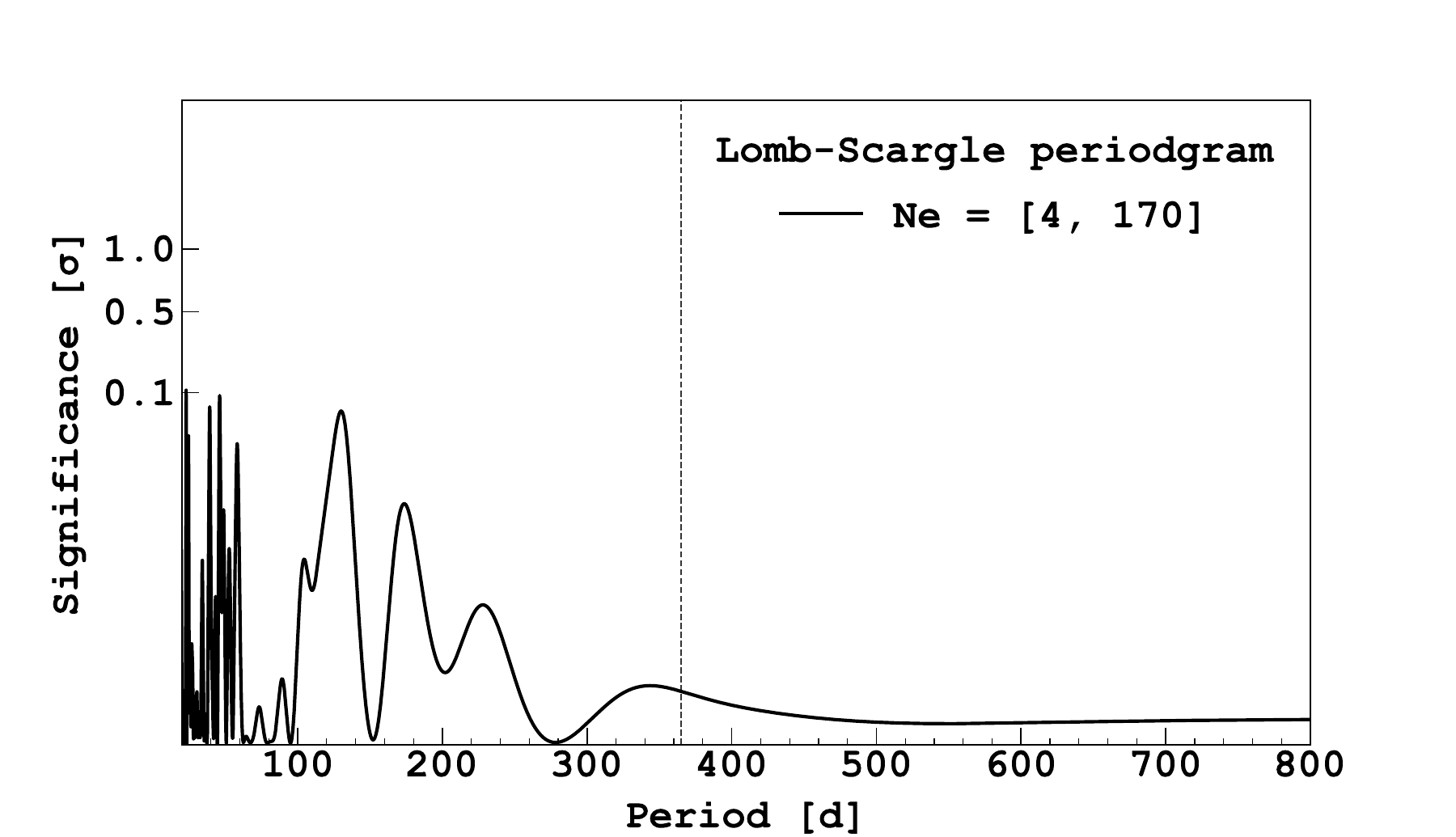}
    \caption{Observed frequency spectrum for the event in~\SIrange{4}{170}{\el} calculated by the Lomb-Scargle periodogram.
    The vertical dotted line corresponds to the frequency of \SI{1}{\yr}.}
    \label{fig:bg_ls}
\end{figure}

Further analyses utilizing both the energy and the temporal information of each event are performed.
The first four months of data are removed from these analyses to avoid over-constraining on the parameters related to the \isotope{37}{Ar} component and the detector calibration.
\reffiginitpar{fig:bgfit} shows the temporal evolution of the event rate in~\SIrange{4}{170}{\el}, after imposing the same fiducialization and event selection criteria as in~\refcite{DarkSide-50:2022qzh}.
The data is fitted with the background model as discussed in the previous section as shown in~\reffig{fig:bgfit}.
The fit returns a \(\chi^2/\mathrm{ndf}\) of 144.0/127.
We then perform the LS analysis on the residual data after subtracting the best-fitted model to detect any sinusoidal event rate modulation over the background.
We do not observe significant periodicity around \SI{1}{\year} as shown in \reffig{fig:bg_ls}.

It is worth noting that in order to exploit, model dependently, the expected signal annual modulation the likelihood of the dark matter spectrum searches in \refscite{DarkSide-50:2022qzh,DarkSide:2022dhx,DarkSide:2022knj} can be extended with an additional term taking into account the timing information of each event.
However, we have verified that, because only a fraction of the dark matter signal would be annually modulated, such combined likelihood does not bring a significant gain in sensitivity, in a realistic situation even in the presence of a hint of a signal emerging from the expected background.

\section{\label{sec:concl}Conclusion}
The \DSf\ dark matter experiment operated an argon TPC filled with UAr for nearly three years.
This paper describes the stability of the detector performance using both the TPC data and various sensors incorporated inside the system.
In particular, the electroluminescence detection efficiency \(g_2\) is confirmed to be stable within fluctuations of no more than~0.5\%, owing to the successful control of the cryogenic system.
Thanks to this stability, we find that both the energy and temporal distribution in \DSf\ are consistent to the background prediction based on the radioisotopes' contamination inside the detector.
A comprehensive study of the detector stability shown here proves that it is possible to further analyze the \DSf\ data with its temporal information~\cite{DarkSide-50:2023fgf}.
In addition, future argon-based dark matter experiments will benefit from the knowledge acquired from the long-term operation of this detector.

\acknowledgments
The DarkSide Collaboration offers its profound gratitude to the LNGS and its staff for their invaluable technical and logistical support. We also thank the Fermilab Particle Physics, Scientific, and Core Computing Divisions. Construction and operation of the DarkSide-50 detector was supported by the U.S. National Science Foundation (NSF) (Grants No. PHY-0919363, No. PHY-1004072, No. PHY-1004054, No. PHY-1242585, No. PHY-1314483, No. PHY-1314501, No. PHY-1314507, No. PHY-1352795, No. PHY-1622415, and associated collaborative grants No. PHY-1211308 and No. PHY-1455351), the Italian Istituto Nazionale di Fisica Nucleare, the U.S. Department of Energy (Contracts No. DE-FG02-91ER40671, No. DEAC02-07CH11359, and No. DE-AC05-76RL01830), the Polish NCN (Grant No. UMO-2019/33/B/ST2/02884) and the Polish Ministry for Education and Science (Grant No. 6811/IA/SP/2018). We also acknowledge financial support from the French Institut National de Physique Nucl\'eaire et de Physique des Particules (IN2P3),   the  IN2P3-COPIN consortium (Grant No. 20-152),  and the UnivEarthS LabEx program (Grants No. ANR-10-LABX-0023 and No. ANR-18-IDEX-0001),  from the São Paulo Research Foundation (FAPESP) (Grants No. 2016/09084-0 and No. 2021/11489-7),  from the Interdisciplinary Scientific and Educational School of Moscow University ``Fundamental and Applied Space Research'',  from the Program of the Ministry of Education and Science of the  Russian  Federation  for  higher  education  establishments,  project No. FZWG-2020-0032 (2019-1569), the International Research Agenda Programme AstroCeNT (MAB/2018/7) funded by the Foundation for Polish Science (FNP) from the European Regional Development Fund, and the European Union's Horizon 2020 research and innovation program under grant agreement No 952480 (DarkWave), and from the Science and Technology Facilities Council, United Kingdom.  I.~Albuquerque is partially supported by the Brazilian Research Council (CNPq). The theoretical calculation of beta decays was performed as part of the EMPIR Project 20FUN04 PrimA-LTD. This project has received funding from the EMPIR program co-financed by the Participating States and from the European Union’s Horizon 2020 research and innovation program. Isotopes used in this research were supplied by the United States Department of Energy Office of Science by the Isotope Program in the Office of Nuclear Physics.


\bibliographystyle{JHEP}
\bibliography{ref.bib}

\providecommand{\href}[2]{#2}\begingroup\raggedright\begin{thebibliography}{10}

\bibitem{Drukier:1986tm}
A.K.~Drukier, K.~Freese and D.N.~Spergel, \emph{{Detecting Cold Dark Matter Candidates}}, \href{https://doi.org/10.1103/PhysRevD.33.3495}{\emph{Phys. Rev. D} {\bfseries 33} (1986) 3495}.

\bibitem{Bernabei:2013xsa}
R.~Bernabei et~al., \emph{{Final model independent result of DAMA/LIBRA-phase1}}, \href{https://doi.org/10.1140/epjc/s10052-013-2648-7}{\emph{Eur. Phys. J. C} {\bfseries 73} (2013) 2648} [\href{https://arxiv.org/abs/1308.5109}{{\ttfamily 1308.5109}}].

\bibitem{Bernabei:2021kdo}
R.~Bernabei et~al., \emph{{Further results from DAMA/Libra-phase2 and perspectives}}, \href{https://doi.org/10.15407/jnpae2021.04.329}{\emph{Nucl. Phys. Atom. Energy} {\bfseries 22} (2021) 329}.

\bibitem{XENON:2017nik}
{\scshape XENON} collaboration, \emph{{Search for Electronic Recoil Event Rate Modulation with 4 Years of XENON100 Data}}, \href{https://doi.org/10.1103/PhysRevLett.118.101101}{\emph{Phys. Rev. Lett.} {\bfseries 118} (2017) 101101} [\href{https://arxiv.org/abs/1701.00769}{{\ttfamily 1701.00769}}].

\bibitem{LUX:2018xvj}
{\scshape LUX} collaboration, \emph{{Search for annual and diurnal rate modulations in the LUX experiment}}, \href{https://doi.org/10.1103/PhysRevD.98.062005}{\emph{Phys. Rev. D} {\bfseries 98} (2018) 062005} [\href{https://arxiv.org/abs/1807.07113}{{\ttfamily 1807.07113}}].

\bibitem{XMASS:2022tkr}
{\scshape XMASS} collaboration, \emph{{Direct dark matter searches with the full data set of XMASS-I}}, {\emph{arXiv} (2022) } [\href{https://arxiv.org/abs/2211.06204}{{\ttfamily 2211.06204}}].

\bibitem{LUX:2016ggv}
{\scshape LUX} collaboration, \emph{{Results from a search for dark matter in the complete LUX exposure}}, \href{https://doi.org/10.1103/PhysRevLett.118.021303}{\emph{Phys. Rev. Lett.} {\bfseries 118} (2017) 021303} [\href{https://arxiv.org/abs/1608.07648}{{\ttfamily 1608.07648}}].

\bibitem{SuperCDMS:2017mbc}
{\scshape SuperCDMS} collaboration, \emph{{Results from the Super Cryogenic Dark Matter Search Experiment at Soudan}}, \href{https://doi.org/10.1103/PhysRevLett.120.061802}{\emph{Phys. Rev. Lett.} {\bfseries 120} (2018) 061802} [\href{https://arxiv.org/abs/1708.08869}{{\ttfamily 1708.08869}}].

\bibitem{XENON:2018voc}
{\scshape XENON} collaboration, \emph{{Dark Matter Search Results from a One Ton-Year Exposure of XENON1T}}, \href{https://doi.org/10.1103/PhysRevLett.121.111302}{\emph{Phys. Rev. Lett.} {\bfseries 121} (2018) 111302} [\href{https://arxiv.org/abs/1805.12562}{{\ttfamily 1805.12562}}].

\bibitem{DarkSide:2018kuk}
{\scshape DarkSide} collaboration, \emph{{DarkSide-50 532-day Dark Matter Search with Low-Radioactivity Argon}}, \href{https://doi.org/10.1103/PhysRevD.98.102006}{\emph{Phys. Rev. D} {\bfseries 98} (2018) 102006} [\href{https://arxiv.org/abs/1802.07198}{{\ttfamily 1802.07198}}].

\bibitem{DEAP:2019yzn}
{\scshape DEAP} collaboration, \emph{{Search for dark matter with a 231-day exposure of liquid argon using DEAP-3600 at SNOLAB}}, \href{https://doi.org/10.1103/PhysRevD.100.022004}{\emph{Phys. Rev. D} {\bfseries 100} (2019) 022004} [\href{https://arxiv.org/abs/1902.04048}{{\ttfamily 1902.04048}}].

\bibitem{COSINE-100:2021xqn}
{\scshape COSINE-100} collaboration, \emph{{Strong constraints from COSINE-100 on the DAMA dark matter results using the same sodium iodide target}}, \href{https://doi.org/10.1126/sciadv.abk2699}{\emph{Sci. Adv.} {\bfseries 7} (2021) abk2699} [\href{https://arxiv.org/abs/2104.03537}{{\ttfamily 2104.03537}}].

\bibitem{LZ:2022ufs}
{\scshape LZ} collaboration, \emph{{First Dark Matter Search Results from the LUX-ZEPLIN (LZ) Experiment}}, {\emph{arXiv} (2022) } [\href{https://arxiv.org/abs/2207.03764}{{\ttfamily 2207.03764}}].

\bibitem{XENON:2023sxq}
{\scshape XENON} collaboration, \emph{{First Dark Matter Search with Nuclear Recoils from the XENONnT Experiment}}, {\emph{arXiv} (2023) } [\href{https://arxiv.org/abs/2303.14729}{{\ttfamily 2303.14729}}].

\bibitem{Amare:2021yyu}
J.~Amare et~al., \emph{{Annual modulation results from three-year exposure of ANAIS-112}}, \href{https://doi.org/10.1103/PhysRevD.103.102005}{\emph{Phys. Rev. D} {\bfseries 103} (2021) 102005} [\href{https://arxiv.org/abs/2103.01175}{{\ttfamily 2103.01175}}].

\bibitem{COSINE-100:2021zqh}
{\scshape COSINE-100} collaboration, \emph{{Three-year annual modulation search with COSINE-100}}, \href{https://doi.org/10.1103/PhysRevD.106.052005}{\emph{Phys. Rev. D} {\bfseries 106} (2022) 052005} [\href{https://arxiv.org/abs/2111.08863}{{\ttfamily 2111.08863}}].

\bibitem{COSINE-100:2022dvc}
{\scshape COSINE-100} collaboration, \emph{{An induced annual modulation signature in COSINE-100 data by DAMA/LIBRA\textquoteright{}s analysis method}}, \href{https://doi.org/10.1038/s41598-023-31688-4}{\emph{Sci. Rep.} {\bfseries 13} (2023) 4676} [\href{https://arxiv.org/abs/2208.05158}{{\ttfamily 2208.05158}}].

\bibitem{DarkSide-50:2022qzh}
{\scshape DarkSide-50} collaboration, \emph{{Search for low-mass dark matter WIMPs with 12~ton-day exposure of DarkSide-50}}, \href{https://doi.org/10.1103/PhysRevD.107.063001}{\emph{Phys. Rev. D} {\bfseries 107} (2023) 063001} [\href{https://arxiv.org/abs/2207.11966}{{\ttfamily 2207.11966}}].

\bibitem{XENON:2016jmt}
{\scshape XENON} collaboration, \emph{{Low-mass dark matter search using ionization signals in XENON100}}, \href{https://doi.org/10.1103/PhysRevD.94.092001}{\emph{Phys. Rev. D} {\bfseries 94} (2016) 092001} [\href{https://arxiv.org/abs/1605.06262}{{\ttfamily 1605.06262}}].

\bibitem{XENON:2019gfn}
{\scshape XENON} collaboration, \emph{{Light Dark Matter Search with Ionization Signals in XENON1T}}, \href{https://doi.org/10.1103/PhysRevLett.123.251801}{\emph{Phys. Rev. Lett.} {\bfseries 123} (2019) 251801} [\href{https://arxiv.org/abs/1907.11485}{{\ttfamily 1907.11485}}].

\bibitem{DarkSide:2015cqb}
{\scshape DarkSide} collaboration, \emph{{Results From the First Use of Low Radioactivity Argon in a Dark Matter Search}}, \href{https://doi.org/10.1103/PhysRevD.93.081101}{\emph{Phys. Rev. D} {\bfseries 93} (2016) 081101} [\href{https://arxiv.org/abs/1510.00702}{{\ttfamily 1510.00702}}].

\bibitem{DarkSide:2018bpj}
{\scshape DarkSide} collaboration, \emph{{Low-Mass Dark Matter Search with the DarkSide-50 Experiment}}, \href{https://doi.org/10.1103/PhysRevLett.121.081307}{\emph{Phys. Rev. Lett.} {\bfseries 121} (2018) 081307} [\href{https://arxiv.org/abs/1802.06994}{{\ttfamily 1802.06994}}].

\bibitem{DarkSide:2018ppu}
{\scshape DarkSide} collaboration, \emph{{Constraints on Sub-GeV Dark-Matter\textendash{}Electron Scattering from the DarkSide-50 Experiment}}, \href{https://doi.org/10.1103/PhysRevLett.121.111303}{\emph{Phys. Rev. Lett.} {\bfseries 121} (2018) 111303} [\href{https://arxiv.org/abs/1802.06998}{{\ttfamily 1802.06998}}].

\bibitem{DarkSide:2022dhx}
{\scshape DarkSide} collaboration, \emph{{Search for Dark-Matter\textendash{}Nucleon Interactions via Migdal Effect with DarkSide-50}}, \href{https://doi.org/10.1103/PhysRevLett.130.101001}{\emph{Phys. Rev. Lett.} {\bfseries 130} (2023) 101001} [\href{https://arxiv.org/abs/2207.11967}{{\ttfamily 2207.11967}}].

\bibitem{DarkSide:2022knj}
{\scshape DarkSide} collaboration, \emph{{Search for Dark Matter Particle Interactions with Electron Final States with DarkSide-50}}, \href{https://doi.org/10.1103/PhysRevLett.130.101002}{\emph{Phys. Rev. Lett.} {\bfseries 130} (2023) 101002} [\href{https://arxiv.org/abs/2207.11968}{{\ttfamily 2207.11968}}].

\bibitem{Back:2012pg}
H.O.~Back et~al., \emph{{First Large Scale Production of Low Radioactivity Argon From Underground Sources}}, {\emph{arXiv} (2012) } [\href{https://arxiv.org/abs/1204.6024}{{\ttfamily 1204.6024}}].

\bibitem{DarkSide:2012fps}
{\scshape DarkSide} collaboration, \emph{{First Commissioning of a Cryogenic Distillation Column for Low Radioactivity Underground Argon}}, {\emph{arXiv} (2012) } [\href{https://arxiv.org/abs/1204.6061}{{\ttfamily 1204.6061}}].

\bibitem{Xu:2012ffp}
J.~Xu et~al., \emph{{A Study of the Residual $^{39}Ar$ Content in Argon from Underground Sources}}, \href{https://doi.org/10.1016/j.astropartphys.2015.01.002}{\emph{Astropart. Phys.} {\bfseries 66} (2015) 53} [\href{https://arxiv.org/abs/1204.6011}{{\ttfamily 1204.6011}}].

\bibitem{DarkSide:2017odo}
{\scshape DarkSide} collaboration, \emph{{The Electronics, Trigger and Data Acquisition System for the Liquid Argon Time Projection Chamber of the DarkSide-50 Search for Dark Matter}}, \href{https://doi.org/10.1088/1748-0221/12/12/P12011}{\emph{JINST} {\bfseries 12} (2017) P12011} [\href{https://arxiv.org/abs/1707.09889}{{\ttfamily 1707.09889}}].

\bibitem{saes}
``Saes getters.'' \url{www.saesgetters.com}.

\bibitem{DarkSide:2021bnz}
{\scshape DarkSide} collaboration, \emph{{Calibration of the liquid argon ionization response to low energy electronic and nuclear recoils with DarkSide-50}}, \href{https://doi.org/10.1103/PhysRevD.104.082005}{\emph{Phys. Rev. D} {\bfseries 104} (2021) 082005} [\href{https://arxiv.org/abs/2107.08087}{{\ttfamily 2107.08087}}].

\bibitem{DarkSide-50:2023fgf}
{\scshape DarkSide-50} collaboration, \emph{{Search for dark matter annual modulation with DarkSide-50}},  \href{https://arxiv.org/abs/2307.07249}{{\ttfamily 2307.07249}}.

\bibitem{DarkSide:2014llq}
{\scshape DarkSide} collaboration, \emph{{First Results from the DarkSide-50 Dark Matter Experiment at Laboratori Nazionali del Gran Sasso}}, \href{https://doi.org/10.1016/j.physletb.2015.03.012}{\emph{Phys. Lett. B} {\bfseries 743} (2015) 456} [\href{https://arxiv.org/abs/1410.0653}{{\ttfamily 1410.0653}}].

\bibitem{Doke:2002oab}
T.~Doke, A.~Hitachi, J.~Kikuchi, K.~Masuda, H.~Okada and E.~Shibamura, \emph{{Absolute Scintillation Yields in Liquid Argon and Xenon for Various Particles}}, \href{https://doi.org/10.1143/JJAP.41.1538}{\emph{Jap. J. Appl. Phys.} {\bfseries 41} (2002) 1538}.

\bibitem{Lomb:1976}
N.R.~{Lomb}, \emph{{Least-Squares Frequency Analysis of Unequally Spaced Data}}, \href{https://doi.org/10.1007/BF00648343}{\emph{\apss} {\bfseries 39} (1976) 447}.

\bibitem{Scargle:1982}
J.~Scargle, \emph{Studies in astronomical time series analysis. ii - statistical aspects of spectral analysis of unevenly spaced data}, \href{https://doi.org/10.1086/160554}{\emph{The Astrophysical Journal} {\bfseries 263} (1983) }.

\bibitem{2018ApJS..236...16V}
J.T.~{VanderPlas}, \emph{{Understanding the Lomb-Scargle Periodogram}}, \href{https://doi.org/10.3847/1538-4365/aab766}{\emph{\apjs} {\bfseries 236} (2018) 16} [\href{https://arxiv.org/abs/1703.09824}{{\ttfamily 1703.09824}}].

\bibitem{DarkSide:2016ddo}
{\scshape DarkSide} collaboration, \emph{{Effect of Low Electric Fields on Alpha Scintillation Light Yield in Liquid Argon}}, \href{https://doi.org/10.1088/1748-0221/12/01/P01021}{\emph{JINST} {\bfseries 12} (2017) P01021} [\href{https://arxiv.org/abs/1611.00241}{{\ttfamily 1611.00241}}].

\bibitem{Monteiro:2008zz}
C.M.B.~Monteiro, J.A.M.~Lopes, J.F.C.A.~Veloso and J.M.F.~dos Santos, \emph{{Secondary scintillation yield in pure argon}}, \href{https://doi.org/10.1016/j.physletb.2008.08.030}{\emph{Phys. Lett. B} {\bfseries 668} (2008) 167}.

\bibitem{WArP:2008dyo}
{\scshape WArP} collaboration, \emph{{Oxygen contamination in liquid Argon: Combined effects on ionization electron charge and scintillation light}}, \href{https://doi.org/10.1088/1748-0221/5/05/P05003}{\emph{JINST} {\bfseries 5} (2010) P05003} [\href{https://arxiv.org/abs/0804.1222}{{\ttfamily 0804.1222}}].

\bibitem{DarkSide:2017wdu}
{\scshape DarkSide} collaboration, \emph{{Simulation of argon response and light detection in the DarkSide-50 dual phase TPC}}, \href{https://doi.org/10.1088/1748-0221/12/10/P10015}{\emph{JINST} {\bfseries 12} (2017) P10015} [\href{https://arxiv.org/abs/1707.05630}{{\ttfamily 1707.05630}}].

\end{thebibliography}\endgroup






\end{document}